\documentclass{ws-procs961x669}            

\newcommand{\be}{\begin{equation}}
\newcommand{\ee}{\end{equation}}
\newcommand{\bea}{\begin{eqnarray}}
\newcommand{\eea}{\end{eqnarray}}
\newcommand{\ba}{\begin{array}}
\newcommand{\ea}{\end{array}}
\newcommand{\ben}{\begin{enumerate}}
\newcommand{\een}{\end{enumerate}}
\newcommand{\ei}{\end{itemize}}
\newcommand{\bc}{\begin{center}}
\newcommand{\bfig}{\begin{figure}}
\newcommand{\efig}{\end{figure}}
\newcommand{\bq}{\begin{quotation}}
\newcommand{\eq}{\end{quotation}}
\newcommand{\bt}{\begin{table}}
\newcommand{\et}{\end{table}}
\newcommand{\btab}{\begin{tabular}}
\newcommand{\etab}{\end{tabular}}
\newcommand{\bs}{\begin{slide}}
\newcommand{\es}{\end{slide}}

\newcommand{\pa}{\partial}

\newcommand{\IR}{\mathbb{R}}

\newcommand{\X}{\mathbb{X}}

\newcommand{\vev}[1]{\langle #1 \rangle}

\newcommand{\red}[1]{{\color{red} #1}}
\newcommand{\blue}[1]{{\color{blue} #1}}

\newcommand{\cyan}[1]{{\color{cyan} #1}}

\def\pa{\partial}
\def\om{\omega}

\def\S{\mathbb{S}}
\def\s{\sigma}

\def\S{\mathbb{S}}

\def\S{\mathbb{S}}

\newcommand{\K}{\mathbb{K}}

\def\pa{\partial}
\def\om{\omega}

\def\S{\mathbb{S}}
\def\s{\sigma}

\def\Ph{\cal{P}}

\def\S{\mathbb{S}}
\def\s{\sigma}

\def\Ph{{\cal{P}}}

\def\tx{\tilde{x}}

\def\tk{\tilde{k}}

\newcommand{\beq}{\begin{eqnarray}}
\newcommand{\eeq}{\end{eqnarray}}

\let\ba=\overline

\let\L=\Lambda

\def\vev#1{\left\langle#1\right\rangle}

\def\tx{\mathord{\tilde x}}
\def\htx{\mathord{\hat{\tilde x}}}

\let\w=\omega

\def\IR{\relax\leavevmode{\rm I\kern-.18em R}}
\def\ZZ{\relax\leavevmode
       \ifmmode\mathchoice
       {\hbox{\sf Z\kern-.4em Z}}
       {\hbox{\sf Z\kern-.4em Z}}
       {\lower.9pt\hbox{\scriptsize\sf Z\kern-.36em Z}}
       {\lower1.2pt\hbox{\tiny\sf Z\kern-.36em Z}}
       \else{\sf Z\kern-.4em Z}\fi}
\def\RR{\relax\leavevmode
       \ifmmode\mathchoice
       {\hbox{\sf R\kern-.4em R}}
       {\hbox{\sf R\kern-.4em R}}
       {\lower.9pt\hbox{\scriptsize\sf R\kern-.36em R}}
       {\lower1.2pt\hbox{\tiny\sf R\kern-.36em R}}
       \else{\sf R\kern-.4em R}\fi}

\def\resetby#1#2{\@addtoreset{#2}{#1}}
\def\seceq{\@addtoreset{equation}{section}
              \def\theequation{\thesection.\arabic{equation}}}

\def\Label#1{\label{#1}%
                \smash{\hbox to0pt{\raise1ex\hbox{\tiny[#1]}\hss}}}
\def\noLabels{\let\Label=\label}

\def\TeV{\text{T\kern0pte\kern-1ptV}}

\begin{document}
\title{On Quantum Gravity and Quantum Gravity Phenomenology
}

\author{Douglas Edmonds}
\address{Department of Physics, Penn State Hazleton,
Hazleton, PA, 18202, U.S.A.\\
E-mail: bde12@psu.edu}

\author{Djordje Minic$^*$ and Tatsu Takeuchi}
\address{Department of Physics, Virginia Tech,
Blacksburg, VA 24061, U.S.A.\\
\smallskip
$^*$Talk given by D.M. at the 16th Marcel Grossmann Meeting\\ 
E-mail: dminic@vt.edu, takeuchi@vt.edu\\
}

\begin{abstract}
This article summarizes a new approach to quantum gravity based on the concepts of
modular spacetime, Born geometry, and metastring theory and their applications to
quantum gravity phenomenology. In particular, we discuss a new understanding of dark matter in terms of metaparticles (zero modes of the metastring) and its relation to dark energy (the curvature of dual spacetime) in view of the actual astronomical observations.
\end{abstract}

\keywords{Quantum Gravity, Quantum Spacetime, Dark Matter, Dark Energy.}

\bodymatter
\section{Introduction and Summary}

In this paper we review recent work 
\cite{Freidel:2013zga, Freidel:2014qna, Freidel:2015pka, Freidel:2015uug, Freidel:2016pls, Freidel:2017xsi, Freidel:2017wst, Freidel:2017nhg, Freidel:2018apz, Freidel:2019jor}
on quantum foundations of quantum mechanics (QM), quantum field theory (QFT),
and quantum gravity (in the guise of metastring theory) as well as unique implications for the problems
of dark matter and dark energy.\cite{rBHM8, Ho:2010ca}
This generic formulation of quantum gravity implies a radiatively stable positive cosmological constant  (dark energy) \cite{rBHM8} in
the observed classical spacetime, and metaparticle quanta (the zero modes of the metastring) 
representing the natural quanta of dark matter \cite{Ho:2010ca} (correlated to dark energy
and visible matter).
The logic of our story is very similar to the path that leads from the Minkowski geometry of special relativity via relativistic non-gravitational field theory
to a dynamical spacetime of general relativity.
In this paper, we start with a hidden geometry in quantum theory (Born geometry) 
and proceed to its dynamical implementation
in quantum gravity (formulated as a metastring theory) with implications for QFT 
(formulated in a way that takes into account the hidden Born geometry) with implications for the
observed world: metaparticles as dark matter quanta, and dark energy emerging from the geometry of
the dual spacetime.
In some sense, this is a sharpening of the modern approaches to non-perturbative quantum physics 
\cite{Wilson:1973jj}, using a simple but crucial insight about a completeness of quantum kinematics 
of discretized physical systems.\cite{Zak:1968zz}


In particular, in Ref.~\citenum{Freidel:2016pls} we demonstrated that any quantum theory is endowed with
a generic quantum polarization associated with modular spacetime.\cite{Freidel:2015uug}
The generic polarization manifestly realizes quantum non-locality, associated with the quantum superposition principle,
that is consistent with causality, and reveals a novel geometry structure, called Born geometry 
\cite{Freidel:2013zga, Freidel:2014qna}, which unifies symplectic ($\omega$), 
orthogonal ($\eta$), and conformal geometries ($H$).
Born geometry is fundamental for a particular quantum theory that consistently propagates
in this geometry -- this turns out to be string theory formulated in a generalized-geometric and intrinsically non-commutative, doubled, chiral phase-space-like form (called metastring theory).\cite{Freidel:2015pka,  Freidel:2017xsi, Freidel:2017wst, Freidel:2017nhg}
The zero modes of the metastring correspond to metaparticles that explicitly realize the
geometry of modular spacetime,\cite{Freidel:2018apz} and as such, could be considered as an explicit
prediction of the modular representation of quantum theory. 
The metaparticles\cite{Freidel:2019jor} are quanta of a modular
generalization of quantum fields, the low energy remnants of metastring fields.
From this new viewpoint\cite{Freidel:2013zga, Freidel:2014qna, Freidel:2015pka, Freidel:2015uug, Freidel:2016pls, Freidel:2017xsi, Freidel:2017wst, Freidel:2017nhg, Freidel:2018apz, Freidel:2019jor} 
quantum gravity is essentially defined as ``gravitization of the quantum,'' that is, as a theory of
a dynamical Born geometry. 
As such it incorporates the concept of Born reciprocity \cite{born} as a covariant implementation of T-duality, the fundamental relation between short and long distance physics in string theory, as well as the new idea of relative (or observer dependent) locality.\cite{AmelinoCamelia:2011bm}

In what follows, we start with a discussion of quantum theory via quantum (modular) spacetime and then comment on QFT in this approach, and then move on to quantum gravity and its phenomenology in the context of dark matter and dark energy and the actual astronomical observations.

\section{Quantum theory and quantum spacetime}

The fundamental reason for the existence of modular polarizations in quantum theory 
is as follows.\cite{Freidel:2016pls}
If one imagines that a quantum system is formulated on a lattice, as assumed in the
modern (Wilsonian) non-perturbative approaches,\cite{Wilson:1973jj} then a theorem due to Zak \cite{Zak:1968zz}
states that a complete set of quantum numbers needed to describe any quantum system would require both
quantum numbers associated with the lattice and its inverse. 
This is easy to see by realizing that non-commuting Hermitian operators, such as coordinates and momenta
$[\,\hat{q},\, \hat{p}\,] = i \hbar$, when exponentiated, together with the appropriate lattice spacing 
$a$ and its inverse ${2\pi \hbar}/{a}$,
commute, that is, 
\begin{equation}
\bigg[
\exp\bigg(\frac{i}{\hbar}\hat{q} \frac{2 \pi \hbar}{a}\bigg),\, 
\exp\bigg(\frac{i}{\hbar}\hat{p} a\bigg)
\bigg]\,=\,0\;.
\end{equation}
Such unitary observables were labeled as ``modular'' by Aharonov.\cite{Aharonov:1969qx} 
These variables are purely quantum in the sense that their formal $\hbar \to 0$ limit is singular.
Also, even though their commutators are zero, the associated Poisson brackets are non-zero, as these are unitary (phase)
variables. Finally, the classical limit is defined by starting with a modular formulation and defining an
appropriate ``extensification," \cite{Freidel:2016pls} with, in principle, many classical limits.
These purely quantum variables also appear in the context of 
QFT.

Thus we concentrate on a complete set of unitary operators as opposed to a complete
set of Hermitian operators. Let us examine the simplest example of the 
$\hat{q}$ and $\hat{p}$ operators. 
The commuting subalgebra of the original non-commuting $[\,\hat{q},\,\hat{p}\,] = i \hbar$ algebra can be completely described by self-dual lattices (endowed with the natural symplectic 
form ($\omega$) coming from the commutator bracket). 
These in turn represent a discretization of a (covariant) phase space defined by $q$ and $p$, 
and when lifted to the original non-commutative algebra, require extra data associated with the lift 
that is described by a doubly orthogonal ($O(d,d)$, where $d$ denotes the spacetime dimension) 
metric $\eta$ (a symmetric counterpart of the antisymmetric $\omega$ associated with $Sp(2d)$ transformations).
Finally, in order to define the vacuum state on this self-dual lattice, we need a conformal structure
$O(2, 2(d-1))$.\cite{Freidel:2016pls} 
This triplet of structures defines Born geometry\cite{Freidel:2013zga, Freidel:2014qna} associated with the modular representation of quantum theory.\cite{Freidel:2016pls} 
Born geometry captures quantum non-locality that is consistent with causality, given
the quantum nature of the unitary operators and the fact that the triple intersection of $Sp(2d)$, $O(d,d)$
and $O(2, 2(d-1))$ gives the Lorentz group.\cite{Freidel:2017xsi}

Let us formalize these insights about the hidden quantum spacetime geometry of quantization.\cite{Freidel:2016pls}
We start with the Heisenberg (or Weyl-Heisenberg) group, which is generated, on the level of the corresponding
algebra, by the familiar position $\hat{q}^a$ and momentum $\hat{p}_b$ operators:
$[\, \hat{q}^a,\, \hat{p}_b\,] = i \hbar \delta^a_b$ .
It will be convenient to introduce a length scale $\lambda$ and a momentum scale $\epsilon$, 
with $\lambda \epsilon = \hbar$.
Also, let us introduce the notation
$\hat{x}^a \equiv \hat{q}^a/\lambda$, $\hat{\tilde{x}}_a \equiv \hat{p}_a/\epsilon$ 
so that $ [\, \hat{x}^a,\, \hat{\tilde{x}}_b\,] = i \delta^a_b$ ,
and let us define 
\begin{equation}
\X^{A}\equiv (x^a , \tilde{x}_a)^{T}\,, \;\; [\, \hat{\X}^a,\, \hat{\X}^b\,] = i \omega^{AB}\,,
\quad\text{with}\quad 
\frac{1}{2}\omega_{AB} d X^A d X^B = \frac{1}{\hbar} dp_{a} \wedge dq^{a}, 
\end{equation}
where 
$\omega_{AB} = - \omega_{BA}$ is the canonical symplectic form on phase space $\cal{P}$.
The Heisenberg group $H_{\cal{P}}$ is generated by Weyl operators \cite{weyl}
$W_{\K} \equiv e^{2 \pi i \omega( \K, \X)}$,
where $\K$ stands for the pair $(\tk,k)$ and  $\omega(\K,\K')=k \cdot \tilde{k}' - \tilde{k}\cdot k'$ .
These form a central extension of the translation algebra,
$W_{\K}W_{\K'} = e^{2 \pi i \omega( \K, \K' )} W_{\K + \K'}$ .
The projection $\pi : H_{\cal{P}} \to \cal{P}$ (where $\pi : W_{\K} \to \K$) defines a line bundle over $\cal{P}$ (in principle a covariant phase space of quantum probes).
In this formulation, states are sections of degree one, 
$W_{\K'} \Phi(\K) = e^{2 \pi i \omega( \K, \K' )} \Phi(\K + \K')$.

Using notions of non-commutative algebra and non-commutative geometry \cite{connes} (such as the theorem of Gelfand-Naimark \cite{GN}), we can say that a 
Lagrangian submanifold (a half-dimensional submanifold of phase space upon which the symplectic form pulls back to zero) is a maximally commutative subgroup of the Heisenberg group.
If we accept this notion of a Lagrangian, then the quantum regime is very different from the
classical regime.
In particular the vanishing Poisson bracket $\{f(q),g(p)\}=0$ requires either $f$ or $g$ to be constant.
However, the vanishing commutator $[f(\hat{q}),g(\hat{p})]=0$ requires only that the functions be commensurately
periodic :
\begin{equation}
e^{i \alpha \hat{p}} e^{i \beta \hat{q}} = e^{i \hbar \alpha \beta} e^{i \beta \hat{q}} e^{i \alpha \hat{p}}\;, 
\qquad
\alpha \beta = 2 \pi/\hbar\;.
\end{equation}
Similar considerations led Aharonov to introduce {\it modular variables} to describe purely quantum phenomena
such as interference.\cite{aharonov2008quantum}


Modular variables are described in great detail in Ref.~\citenum{aharonov2008quantum}, where one can find detailed references on this subject.
The modular  variables, denoted $[\,\hat{q}\,]$ and $[\,\hat{p}\,]$, are 
defined modulo a length scale $R$ (the slit spacing being a natural choice) as
\begin{equation}
[\,\hat{p}\,]_{2 \pi \hbar/R} = \hat{p} \;\mathrm{mod}\left(2 \pi \hbar/R\right)\;,\qquad
[\,\hat{q}\,]_R = \hat{q} \;\mathrm{mod}\left( R\right)\;.
\end{equation}
They play a central role in
understanding interference in terms of operators (and not states).
The shift operator $e^{i R\hat{p}/\hbar}= e^{iR [\,\hat{p}\,]/\hbar}$ shifts the position of a particle state (say an electron in the double-slit experiment) by distance $R$ and is a function of the modular momenta (see also Ref.~\citenum{Chang:2012eh}).  
These modular variables (the main examples being the Aharonov-Bohm and Aharonov-Casher phases \cite{aharonov2008quantum}) satisfy non-local operator equations of motion. For example, given the Hamiltonian $\hat{H} = \hat{p}^2/2m +V(\hat{q})$, the 
Heisenberg equation of motion for the shift operator is
\begin{equation}
e^{-i R\hat{p}/\hbar}  \,\frac{d}{dt}\,e^{iR\hat{p}/\hbar} 
\;=\; -\frac{iR }{\hbar}\left(\frac{V(\hat{q}+R)-V(\hat{q})}{R}\right).
\end{equation}
Modular variables are fundamentally
non-local in a non-classical sense, since we see here that their evolution depends on the value of the potential at distinct locations. 
Remarkably, thanks to the uncertainty principle, this dynamical non-locality does not lead to a violation of causality.\cite{aharonov2008quantum}
One of the characteristic features of these variables is that they do not have classical analogues; indeed, the limit $\hbar \to 0$ of $[\,p\,]_{h /R}$ is ill-defined. 
Also, modular variables capture entanglement of continuous $q$, $p$ variables.
When exponentiated (\textit{i.e.}, when understood as Weyl operators) the modular variables naturally commute. 
In other words, given $[\,\hat{x}^a, \,\hat{\tilde{x}}_b\,] = i\delta^a_b$
we find\cite{Freidel:2016pls}
$[\,e^{2\pi i\hat{x}},\, e^{2\pi i \hat{\tilde{x}}}\,] =0$ .
Thus, the quantum algebra of modular variables possesses more commutative directions than the classical Poisson algebra,
since the Poisson bracket of modular variables does not vanish, 
$\{\,e^{2\pi i x}, \, e^{2\pi i \tilde{x}\,} \}\neq 0$ .

\subsection{Modular spacetime as quantum spacetime and Born geometry}

In this view of quantum theory we have a structure analogous to a Brillouin cell 
in condensed matter physics.
The volume and shape of the cell are given by $\lambda$ and $\epsilon$ (i.e. $\hbar$ and $G$ (or $\alpha'$)).
The uncertainty principle  is implemented in a subtle way: we can specify a point in a modular cell, 
but if so, we cannot say {\it which} cell we are in.
This means that there is a more general notion of quantization.\cite{Freidel:2016pls}
Instead of selecting a classical polarization $L$ (the arguments of the wave function, or the arguments of a local quantum field)
we choose a {\it modular polarization}.
In terms of the Heisenberg group, all that is happening is that in order to have a commutative algebra, we need only
$\omega(\K, \K') \in 2\mathbb{Z}$, and $ W_{\K}W_{\K'} = e^{2 \pi i \omega( \K, \K' )} W_{\K + \K'} = W_{\K'}W_{\K'}$.
This defines a lattice $\Lambda$ in phase space $\cal{P}$.
Finally, we specify a ``lift'' of the lattice from the phase space $\cal{P}$ to the Heisenberg group $H_{\cal{P}}$.

Maximally commuting subgroups $\hat{\Lambda}$ of the Heisenberg group correspond to lattices that are integral and self-dual with respect to $\omega$.\cite{Mackey} 
Given $W_{\lambda}$, where $\lambda \in \Lambda$, there is a lift to $\hat{\Lambda}$  which defines ``modular polarization,''
$U_{\lambda} = \alpha(\lambda) W_{\lambda}$,
where $\alpha(\lambda)$ satisfies the co-cycle condition
$ \alpha(\lambda)\, \alpha(\mu)\, e^{ \pi i \omega( \lambda, \mu)} =\alpha(\lambda + \mu)$ ,
with $\lambda, \alpha \in \Lambda $ .
One can parametrize a solution to the co-cycle condition by introducing a symmetric bilinear from 
$\eta(\K,\K') =k \cdot \tilde{k}' + \tilde{k}\cdot k'$
and setting
$\alpha_{\eta}(\lambda) \equiv e^{i \frac{\pi}{2}  \eta( \lambda, \lambda)}$.
Finally, when we choose a classical Lagrangian $L$ there exists a special translation-invariant 
state that we associate with the vacuum, which we interpret as ``empty space.''
In modular quantization there is no such translation-invariant state because of the lattice structure.
The best we can do is to choose a state that minimizes an ``energy,'' which requires the introduction of another
symmetric bilinear form that we call, again suggestively, $H$.
This means that we are looking for 
operators such that
\be\label{Comm1}
[\,\hat{\mathbb{P}}_A,\, \Phi\,]= \frac{i}{2\pi}\partial_A \Phi\;,\quad\text{and}\quad 
\Phi(\hat{\X}+\lambda)=\Phi(\hat\X)\;,
\ee
where the modular observables  $\Phi(\hat{\X}+{\lambda})=\Phi(\X)$ are  generated by the lattice observables $U_\lambda$ with $\lambda \in \Lambda$. 
Translation invariance would be the condition $\hat{\mathbb{P}}\,|0\rangle=0$. Since this is not possible, the next natural choice is to minimize the translational energy. 
Therefore, we pick a positive definite metric $H_{AB}$ on $\Ph$, and we define \cite{Freidel:2016pls}
$\hat{E}_H\equiv H^{AB} \hat{\mathbb{P}}_A\hat{\mathbb{P}}_B$,
and demand that $|0\rangle_H$ be the ground state of $\hat{E}_H$.
This is indeed the most natural choice and it shows that we cannot fully disentangle kinematics (i.e., the definition of translation generators) from dynamics.
In the Schr\"odinger case, since the translation generators commute, the vacuum state $\hat{E}\,|0\rangle=0$ is also the translation-invariant state and it carries no memory of the metric $H$ needed to define the energy.
In our context, due to the non-commutativity of translations, the operators $\hat{E}_H$ and $\hat{E}_{H'}$ do not commute. Thus, the vacuum state depends on $H$, in other words $|0\rangle_H\neq |0\rangle_{H'}$, and it also possesses a non-vanishing zero point energy.

Therefore, modular quantization involves the introduction of three quadratic forms $(\omega, \eta, H)$, called {\it Born geometry},\cite{Freidel:2013zga, Freidel:2014qna} which underlies the geometry of modular variables.
As we will see, in the context of metastring theory, a choice of polarization is a choice of a spacetime within
$\cal{P}$ but the most general choice is a {\it modular polarization} that we have discussed above.
From the foundational quantum viewpoint, Born geometry $(\omega, \eta, H)$ arises as a parametrization of such quantizations, which
results in a notion of quantum spacetime, that we call {\it modular spacetime}. 
In particular, a one-dimensional 
modular line is a two-dimensional 
torus that is compact and not simply connected.
Finally, large spacetimes  of canonical general relativity (and its extensions, like string theory) result as a ``many-body'' phenomenon through a process of tensoring (entangling) unit modular cells, which we refer to as ``extensification.'' \cite{Freidel:2016pls}
Note that the Lorentz group (in $d$ spacetime dimensions) lies at the intersection of the symplectic, neutral and doubly orthogonal groups,\cite{Freidel:2016pls}
\be\label{orthinter}
{\rm O}(1,d-1)\;=\;{\rm Sp}(2d) \cap {\rm O}(d,d) \cap {\rm O}(2,2(d-1))\;,
\ee
which sheds new light on the origin of quantum theory through compatibility of the causal (Lorentz) structure and non-locality 
captured by the discreteness of quantum spacetime. Note that relative (observer-dependent) locality
\cite{AmelinoCamelia:2011bm} is
needed to resolve the apparent contradiction between discreteness of quantum spacetime and Lorentz symmetry.

One can pass from a classical polarization (such as the  Schr\"odinger representation) to a modular polarization 
via the Zak transform.\cite{Zak:1968zz} 
Note that, there is a connection on the line bundle over phase space that has unit flux through a modular cell. (This is very similar to Integer Quantum Hall effect.)
A modular wave function is quasi-periodic
\begin{equation}
\Psi(x+a, \tilde{x}) \,=\, e^{2 i \pi a \tilde{x}} \Psi (x, \tilde{x})\;,\quad 
\Psi(x, \tilde{x} + \tilde{a}) \,=\, \Psi (x, \tilde{x})\;.
\end{equation}
The quasi-periods correspond to the tails of an Aharonov-Bohm \cite{Aharonov:1959fk} potential attached to a unit flux.
In particular, vacuum states must have at least one zero in a cell, which leads to theta functions (the
Zak transforms of Gaussians).

\subsection{A comment on quantum field theory and quantum spacetime}

A few general comments about QFT in the modular form are in order, following the general modular formulation of any quantum theory. 
The modular polarization of QFT reveals new structures and sheds
new light on both the short distance (UV) and long distance (IR) physics of quantum fields, and the
continuum limit of QFTs which is self-dual with respect to the UV and IR properties
(resembling some crucial properties of non-commutative field theories \cite{Douglas:2001ba}). 
In particular, the modular representation of QFT introduces dual ``electric'' and ``magnetic''
variables, which are non-commuting in general. This extends our results in
the context of the 2$d$ conformal field theory formulation of string theory in which the non-commutativity
of such ``electric'' and ``magnetic'' variables has been explicitly demonstrated.\cite{Freidel:2017wst, Freidel:2017nhg}

The general modular representation can be defined in terms of the Zak transform of a Schr\"{o}dinger
representation (\textit{i.e.}, wave functions).
Given a square normalizable wave function $\psi(x)$ belonging to a Hilbert space, 
one defines the modular representation as the lattice Fourier transform (or Zak transform)
\be
\psi_a(x, \tilde{x}) \;\equiv\; \sqrt{a} \sum_n e^{-2 \pi i n \tilde{x}} \psi(a(x +n))\;,
\ee
where $x\equiv q/a$,  $\tilde{x} \equiv p/b$, with $ab = 2 \pi \hbar$.
Note that if $\psi(x)$ is a Gaussian, its Zak transform, the modular $\psi_a (x, \tilde{x})$, 
is given by the doubly-periodic theta function associated with the lattice.
(The inverse Zak transform
\be
\Phi(x +n ) \;\equiv\; \dfrac{1}{\sqrt{a}} \int_0^1 d\tilde{x}\,e^{2 \pi i n \tilde{x}} \Phi_a (a^{-1} x, \tilde{x}) \;,
\ee
illustrates that the usual Schr\"{o}dinger representation is truly  singular, and thus not generic.)

Now if one second quantizes $\psi(x)$, one naturally ends up with a quantum field operator $\phi(x)$.
Similarly, the second quantization of the modular $\psi_a (x, \tilde{x})$ would lead to
a modular quantum field operator $\phi (x, \tilde{x})$,
$\phi(x) \to \phi (x, \tilde{x})$ .
The excitations of such modular quantum fields are non-local in general, and 
will be discussed below as \textit{metaparticles}.
Note that the usual wave functional approach to QFT
defined in terms of functionals $\Psi\big[\phi(x)\big]$ should be now
defined in terms of wave functionals of modular fields $\Psi\big[\phi (x, \tilde{x})\big]$.
However, now we have more freedom
in the general modular polarization.
The dual momenta $p$ and $\tilde{p}$ (to $x$ and $\tilde{x}$ respectively) lead, 
via the canonical minimal-coupling prescription, not only to the
usual fields $\phi$ but also to their duals $\tilde{\phi}$ (see below).
This procedure defines the modular polarization of QFT in terms of the
functional Zak transform of the original wave functional,
$\Psi\big[\phi(x)\big] \to \Psi\big[\phi(x, \tilde{x}), \tilde{\phi}(x, \tilde{x})\big]$.
For example, the Gaussian wave functionals with non-trivial kernels (such as the
ones found in the context of non-trivial interacting theories like 
2+1 and 3+1 dimensional Yang Mills theory
\cite{Feynman:1981ss}) 
would be mapped into functional theta functions.

The non-perturbative formulation is defined in a symmetric, self-dual way with respect to the double RG flows
(as in non-commutative field theory \cite{Douglas:2001ba}) 
with full spacetime covariance, and 
should be important 
not only for quantum non-locality of QFT, but also in the realms of
strong coupling and deep infrared.

\section{Quantum gravity and dynamical quantum (modular) spacetime}

The unexpected outcome of this new view of the foundations of quantum theory is that this fundamental geometry of quantum theory can be realized in the context of metastring theory, in which this Born geometry
(given by $\omega$, $\eta$ and $H$) is ``gravitized'' (i.e. dynamical).
At the classical level, metastring theory\cite{Freidel:2013zga, Freidel:2014qna, Freidel:2015pka, Freidel:2015uug, Freidel:2016pls, Freidel:2017xsi, Freidel:2017wst, Freidel:2017nhg, Freidel:2018apz, Freidel:2019jor} can be thought of as a formulation of string theory in which the target space is doubled in such a way that T-duality acts linearly on the coordinates. This doubling means that  momentum and winding modes appear on an equal footing.  
In this formulation, T-duality exchanges the Lagrangian sub-manifold with its image under $J=\eta^{-1}H$.
Classical metastring theory is defined by the action \cite{Freidel:2015pka,Tseytlin:1990va} 
\begin{equation}\label{metastringAction}
\hat{S}=  \frac{1}{4\pi}\int_{\Sigma}d^2\sigma \Big(  \pa_{\tau}{\X}^{A} (\eta_{AB}+\omega_{AB}) (\X) \pa_{\sigma}\X^{B} -  
  \pa_{\sigma}\X^{A}  H_{AB} (\X) \pa_{\sigma}\X^{B}\Big)\,,
\end{equation}
where $\X^A$ are dimensionless coordinates on phase space and the fields $\eta, H,\omega$ are all dynamical (i.e., in general dependent on $\X$)  phase space fields
($\X^{A}\equiv (x^a , \tilde{x}_a)^{T}$). 
In the context of a flat metastring we have 
 constant $\eta_{AB}$,  $H_{AB}$,
 and $\omega_{AB}$ :
\be\label{etaH0}
\eta_{AB} \equiv \left( \begin{array}{cc} 0 & \delta \\ \delta^{T}& 0  \end{array} \right),\quad
H_{AB} \equiv  \left( \begin{array}{cc} h & 0 \\ 0 &  h^{-1}  \end{array} \right),\quad
\om_{AB} = \left( \begin{array}{cc} 0 & \delta \\ -\delta^{T}& 0  \end{array} \right),
\ee
where $\delta^{\mu}_{\nu}$ is the $d$-dimensional identity matrix, and
$h_{\mu\nu}$ is the $d$-dimensional Lorentzian metric.
The Polyakov string \cite {Polchinski:1998rq} is obtained in a singular limit of zero 
$\omega$ after integrating over $\tilde{x}$. (For a phase space structure of 
the canonical string see Ref.~\citenum{Minic:1991rk} and references therein.)

\subsection{Non-commutativity and non-associativity in quantum gravity}

The metastring formulation points to an unexpected fundamental non-commutativity of closed string theory, that we address in what follows.
The string commutation relations\cite{Freidel:2017wst,Freidel:2017nhg} state
$[\X^A(\s),\X^B(\s')]=  2i\lambda^2 \left[ \pi \omega^{AB}-\eta^{AB}\theta(\sigma-\s')\right]$,
where $\theta(\s)$ is the staircase distribution, i.e., a solution of $\theta'(\s)=2\pi\delta(\s)$ ; it is odd and  quasi-periodic with period  $2\pi$.
Following standard practice, all indices are raised and lowered using  $\eta$ and $\eta^{-1}$. 
The momentum density operator is given by 
\begin{equation}
{\mathbb{P}}_A(\s) \;=\; \dfrac{1}{2\pi\alpha'}\,\eta_{AB}\,\pa_\s \X^B(\s)\;,
\end{equation} 
and the previous commutation relation implies that it is conjugate to $\X^A(\s)$. 
The two-form 
$\omega$ appears when one integrates this canonical commutation relation to include the zero-modes, the integration constant being uniquely determined by worldsheet causality. 
Denoting by $(\,\hat\X,\,\hat{\mathbb{P}}\,)$ 
the zero mode components of the string operators $\X(\s)$ and ${\mathbb{P}}(\s)$,  
we have
$\big[\,\hat{{\mathbb{P}}}_A,\,\hat{{\mathbb{P}}}_B\big]=0$, 
$\big[\,\hat{\X}^A,\,\hat{{\mathbb{P}}}_B\big] = i\hbar \delta^A{}_B$, 
and $\big[\,\hat{\X}^A,\,\hat{\X}^B\big] = 2\pi i\lambda^2 \omega^{AB}$.
This is a deformation of the doubled Heisenberg algebra involving the string length $\lambda$ as a deformation parameter.  
Note that under a constant $B$-field transformation 
 $\X=(x^a,\tx_a) \mapsto (x^a, \tx_a + B_{ab} x^b)$, 
 the trivial symplectic form $\omega(\K,\K')=k \cdot \tilde{k}' - \tilde{k}\cdot k'$ is mapped onto
$
\omega(\K,\K')=  k_a  \tk'^a - k'_a \tk^a - 2 B_{ab} \tk^a \tk'^b,
$
and the commutators read
\beq\label{Bcommrel}
\big[\,\hat{x}^a,\,\hat{x}^b\,\big]=0\,,\quad 
\big[\,\hat{x}^a,\,\hat{\tx}_b\,\big]=2\pi i\lambda^2 \delta^a{}_b\,,\quad 
\big[\,\hat{\tx}_a,\,\hat{\tx}_b\,\big]=-4\pi i\lambda^2 B_{ab}\,.
\eeq
We see that the  effect of the $B$-field is to render the dual coordinates non-commutative (and that the $B$-field originates from the symplectic structure $\omega$). 
This implies a new view of the axion in four spacetime dimensions.
The $\beta$-transformation on the other hand corresponds to the map $(x^a,\tx_a)\mapsto (x^a+\beta^{ab}\tx_b,\tx_a)$. 
Equivalently, it has the effect of mapping the symplectic structure to
$\omega(\K,\K')=  k_a  \tk'^a - k'_a \tk^a + 2 \beta^{ab} k_a k'_b$,
and yields commutation relations
 \beq\label{betacommrel}
\big[\,\hat{x}^a,\,\hat{x}^b\,\big]=4\pi i\lambda^2 \beta^{ab}\,,\qquad 
\big[\,\hat{x}^a,\,\hat{\tx}_b\,\big]=2\pi i\lambda^2 \delta^a{}_b\,,\qquad 
\big[\,\hat{\tx}_a,\,\hat{\tx}_b\,\big]=0\,.
\eeq
Dramatically, the coordinates that are usually thought of as the spacetime coordinates have become themselves non-commutative. 
Since this is the result of an $O(d,d)$ transformation it can be thought of in similar terms as the $B$-field; these are related by T-duality. We are familiar with the $B$-field background because we have, in the non-compact case, a fixed notion of locality in the target-space theory. However, in the non-geometric $\beta$-field background, we do not have such a notion of locality but we can access it through T-duality.
However, this background does lead to non-commutative field theory, and
one can place a bound on the minimal length $\lambda$ of $O(10\,\mathrm{TeV})$, 
which is the current high-energy limit for probing the continuum structure of
spacetime. (Similarly, this background can be used to argue for an effective minimal-length extension of commutation relations.\cite{tatsuml} 
For astrophysical probes of the minimal length, see Ref.~\citenum{mike}.)

Note that the dilaton can be understood as coming from the volume of phase space.\cite{Boffo:2019zus}
In general, for varying $B$-backgrounds we encounter non-associativity as well,\cite{nonassoc} 
and the proper closure of such non-commutative and non-associative structure is ensured by the equations of motion.
Here we remark that fundamental non-associativity can be related to the robustness of the Standard Model (SM) gauge group.\cite{Gunaydin:1974fb}
Similarly, fundamental non-commutativity can be related to the underlying non-commutative nature of the SM and its phenomenology.\cite{connes, Aydemir:2013zua}

\subsection{Non-perturbative formulation of quantum gravity}

The metastring offers a new view on the fundamental question of a non-perturbative formulation of quantum gravity. \cite{Freidel:2013zga, Freidel:2014qna, Freidel:2015pka, Freidel:2015uug, Freidel:2016pls, Freidel:2017xsi, Freidel:2017wst, Freidel:2017nhg, Freidel:2018apz, Freidel:2019jor}
The worldsheet can be made modular in our formulation, with the doubling of $\tau$ and $\sigma$, so
that  $\X (\tau, \sigma)$ can  be
in general viewed as an infinite-dimensional matrix (the matrix indices coming from the Fourier components of the
doubles of $\tau$ and $\sigma$). Then the corresponding metastring action reads as
\be
\int \mathrm{Tr} \Big[\, \partial_{\tau} \X^A \partial_{\sigma} \X^B (\omega_{AB} + \eta_{AB}) - 
\partial_{\sigma} \X^A H_{AB} \partial_{\sigma} \X^B \,\Big] d \tau d \sigma \;,
\ee
where the trace is over the matrix indices.
One can associate partonic degrees of freedom with matrix entries.
A non-perturbative quantum gravity follows by replacing the $\sigma$-derivative with a
commutator involving one extra $\X^{26}$:
\be
\partial_{\sigma} \X^A \to \big[\,\X^{26},\,\X^A\big]\;,
\ee
with $A=0,1,2,\cdots,25$.
The resulting matrix-model form of the above metastring action is
\be
\int \mathrm{Tr}\Big( \partial_{\tau} \X^a \big[\X^b, \X^c \big] \eta_{abc}  - 
H_{ac} \big[\X^a, \X^b \big] \big[\X^c, \X^d \big] H_{bd} \Big) d \tau\;,
\ee
with $a,b,c=0,1,2,\cdots,25, 26$,
where the first term is of a Chern-Simons form and the second of the Yang-Mills form.
$\eta_{abc}$ contains both $\omega_{AB}$ and $\eta_{AB}$.
{This defines a non-perturbative quantum gravity viewed as ``a gravitization of the quantum.'' \cite{Freidel:2019jor}

This formulations invokes the IIB matrix model,\cite{Ishibashi:1996xs} which describes $N$ D-instantons (and is by T-duality related to the Matrix model of M-theory~\cite{Banks:1996vh}).
Given our new viewpoint we suggest a new covariant non-commutative matrix-model formulation of string theory as a theory of quantum gravity, by writing in the large $N$ limit $\pa_\s\X^C = [\X, \X^C]$ 
(and similarly for $\pa_{\tau}\X^B$) in terms of commutators of two (one for $\pa_\s\X^C$ and one for $\pa_{\tau}\X^C$) extra $N\,{\times}\,N$ matrix-valued chiral $\X$'s.
Notice that, in general, we do not need an overall trace, and so the action can be viewed as a matrix, rendering the entire non-perturbative formulation as
purely quantum in the sense of the original matrix formulation of QM:
\begin{equation}
\S_{\text{ncF}}\,{=}\,\frac{1}{4\pi} 
\big[{\X}^{a},  {\X}^{b}\big] \big[{\X}^{c},  {\X}^{d}\big]  f_{abcd}\;,
\end{equation}
where instead of $26$ bosonic $\X$ matrices we have $28$, with supersymmetry emerging in 10($+$2) dimensions from this underlying
 bosonic formulation. (This is a non-commutative matrix-model formulation of F-theory.)
By T-duality, the new covariant M-theory matrix model is 
\begin{equation}
 \S_{\text{ncM}}\,{=}\,\frac{1}{4\pi}  
\int_{\tau} 
\Big(\pa_{\tau} \X^i \big[{\X}^{j},  {\X}^{k}\big] g_{ijk} 
- \big[{\X}^{i},  {\X}^{j}\big]\big[{\X}^{k},  {\X}^{l}\big] h_{ijkl}\Big),
\end{equation}
with 27 bosonic $\X$ matrices, with supersymmetry emerging in 11 dimensions. 
In this approach,
holography \cite{tHooft:1993dmi} (such as AdS/CFT \cite{Maldacena:1997re} 
or dS/CFT \cite{Balasubramanian:2002zh}),
which can be viewed as a 
``quantum Jarzynski equality on the space of geometrized RG flows,''\cite{Minic:2010pw}
is emergent in a particular ``extensification'' of quantum spacetime.
The relevant information about
 $\w_{AB}$, $\eta_{AB}$, and $H_{AB}$ is now contained in the new dynamical backgrounds $f_{abcd}$ in F-theory, and $g_{ijk}$ and $h_{ijkl} $ in M-theory. 
{This offers a new formulation of covariant Matrix theory in the M-theory limit,\cite{Minic:1999js}
which is essentially a partonic formulation; strings emerge from partonic constituents in a certain limit. This new matrix formulation is fundamentally bosonic (\textit{i.e}, supersymmetry is emergent only in a specific limit) and thus it is reminiscent of bosonic M-theory.\cite{Horowitz:2000gn}
The relevant backgrounds $g_{ijk}$ and $h_{ijkl} $ should be determined by the matrix RG equations.
Also, there are lessons here for the new concept of ``gravitization of quantum theory'' as well as the idea that dynamical Hilbert spaces, or 2-Hilbert spaces (here represented by matrices), are fundamentally needed
in quantum gravity.\cite{twohilbert}}

This matrix-like formulation should be understood as a general non-perturbative formulation of 
string theory. In this partonic formulation closed strings are collective excitations, in turn constructed from the product of open string fields. The observed classical spacetime emerges as an ``extensification'' \cite{Freidel:2013zga, Freidel:2014qna, Freidel:2015pka, Freidel:2015uug, Freidel:2016pls, Freidel:2017xsi, Freidel:2017wst, Freidel:2017nhg, Freidel:2018apz, Freidel:2019jor} in a particular limit, out of the basic building blocks of quantum spacetime.
Their remnants can be found in the low energy bi-local quantum fields, with bi-local (metaparticle) quanta, to which we now turn.


\section{Quantum gravity, metastrings, metaparticles and dark matter}

The above manifestly T-duality covariant formulation of closed strings (i.e. metastring theory) implies intrinsic non-commutativity of zero-modes. It is thus instructive to formulate a particle-like limit of the metastring 
that we call the {\it metaparticle}.\cite{Freidel:2017wst, Freidel:2017nhg, Freidel:2018apz} 
The theory of metaparticles (the low energy remnants of the metastring, and as such, the low energy remnants (predictions) of quantum gravity) is defined by the following world-line action \cite{Freidel:2017wst, Freidel:2017nhg, Freidel:2018apz} 
\begin{equation}\label{1}
S \equiv \int_0^1 d\tau 
\left[p \dot x +\tilde p \dot{\tilde x} + \lambda^2 p \dot{\tilde p} 
- \frac{N}2\left(p^2 +{\tilde p}^2 + m^2\right) +{\tilde N}\left(p \tilde p - \mu \right)
\right]\,.
\end{equation}
Here the signature $(+,-,\cdots,-)$ and the contraction of indices are implicitly assumed.
At the classical level, theory of metaparticles is a worldline theory with the usual reparameterization invariance and two additional features.\cite{Freidel:2018apz} 
The first new feature is the presence of an additional local symmetry, which from the string point of view corresponds to the completion of worldsheet diffeomorphism invariance. From the particle worldline point of view, 
this symmetry is associated with an additional local constraint. The second new feature is the presence of a non-trivial symplectic form on the metaparticle phase space, also motivated by string theory.\cite{Freidel:2017wst, Freidel:2017nhg} 
Because of its interpretation as a particle model on Born geometry, associated with the modular representation of quantum theory,
the spacetime on which the metaparticle propagates is ambiguous, with different choices related by 
T-duality. 
The attractive feature of this model include worldline causality and unitarity, as well as an explicit mixing of widely separated energy-momentum scales.
The metaparticle propagator follows from the world line path integral defined by the above action 
and it has the following form in momentum space\cite{Freidel:2018apz} :
\begin{equation}\label{doubletramp}
G(p,\tilde p; p_i,\tilde p_i) \;\sim\;
\delta^{(d)}(p-p_{i})\,\delta^{(d)}(\tilde p-\tilde p_{i})\,
\frac{\delta(p\cdot\tilde p-\mu)}{p^2+\tilde p^2+m^2-i\varepsilon}\;.
\end{equation}
The canonical particle propagator is a highly singular $\tilde p \to 0$ (and $\mu \to 0$) limit of this expression.
This propagator also predicts the following dispersion relation (in a particular gauge \cite{Freidel:2018apz}) that
can be tested in various experiments and with various probes:
$E_p^2 + (\mu^2/E_p^2) = {\vec{p}}\,{}^2 + m^2$.
This formulation is fully compatible with Lorentz covariance, and is a direct consequence of the consistency of
quantum theory and a minimal length (and thus Born geometry). In a cosmological 
context, one can use this dispersion relation to put a bound on $\mu$ for the
case of neutrinos, which turns out to be close to the energy scale of dark energy.\cite{jerzy}
In general, for each particle at energy $E$ there exists a dual particle at energy $\mu/E$.
This is in complete analogy of the well-known prediction of antiparticles in the union of special relativity and quantum theory.

Note that the usual particle limit is obtained, at least classically, by taking $\mu \to 0$ and $\tilde p \to 0$.
Given the form of the above Lagrangian, the metaparticle looks like two particles that are entangled through a 
Berry-phase-like $p_\mu\, \dot{\tilde p}^\mu$ factor. The metaparticle is fundamentally non-local, and thus it should not be associated with effective local field theory.
In particular, by looking at the metaparticle constraints
$
p^2 + {\tilde{p}}^2 = m^2$ and $ p \tilde{p} = \mu,
$
we note that the momenta $p$ and $\tilde{p}$ can be, in principle, widely separated.
For example, if $m$ is of the order of the Planck energy, and $\mu$ of the order of $\mathrm{TeV}^2$ 
(which can be considered
a characteristic particle physics scale), then the $p$ can be of the order of the Planck energy, and the $\tilde{p}$ of the vacuum energy scale. 
Thus metaparticle theory can naturally relate widely separated scales, which transcends the usual reasoning based on Wilsonian effective field theory (and should be relevant for the naturalness and hierarchy problems).

We can also discuss the background fields that couple to the
metaparticle quanta.
Following the well-known procedure of introducing background fields in the case of particles, by
shifting the canonical momentum by a gauge field we may try to extend the gauging procedure to the metaparticle counterpart.
There is a possible ambiguity in this gauging which depends on which 
configuration variables one decides to work with.
If one takes $(x,\tx)$ as configuration variables,
one obtains a gauging  which could also be motivated
by the presence of a ``stringy gauge field'' in metastring theory,\cite{Freidel:2015pka}
$p_\mu \to p_\mu+A_\mu(x,\tilde x)$ and $\tilde p^\mu \to \tilde p^\mu+\tilde A^\mu(x,\tilde x)$.
The generic prediction here is the existence of a dual field $\tilde{A}$, which is correlated (entangled), with the original $A$ field.
Thus the entire SM would have a dual SM (which we propose,
describes the dark matter sector).

We expect that the correct field theoretic description of the metaparticle is in terms of the above general non-commutative (modular) field theory $\Phi(x,\tx)$ limit of the metastring.\cite{Freidel:2013zga, Freidel:2014qna, Freidel:2015pka, Freidel:2015uug, Freidel:2016pls, Freidel:2017xsi, Freidel:2017wst, Freidel:2017nhg, Freidel:2018apz, Freidel:2019jor} 
Such an effective non-commutative field theory is similar in spirit to 
Ref.~\citenum{Douglas:2001ba}. 
Also, we note that the concept of metaparticles might be argued from the compatibility of the quantum spacetime
that underlies the generic representations of quantum theory, as discussed in Ref.~\citenum{Freidel:2016pls}, and thus the metaparticle might be as ubiquitous as the concept of antiparticles which is demanded by the compatibility of 
relativity and quantum theory.
The metaparticles also provide a natural route to the problem of dark matter.
To lowest (zeroth) order of the expansion in the non-commutative parameter $\lambda$, 
the effective action for the SM matter Lagrangian ($L_m$) and their duals (that could be interpreted as the dark matter Lagrangian $L_{dm}$) takes the form 
\begin{equation}
S_\mathrm{eff} = - \iint\! \sqrt{{g(x)}{\tilde{g}(\tx)}} \big[ L_m (A(x, \tilde{x})) + \tilde{L}_{dm}   ( \tilde{A}(x, \tilde{x}))+ \cdots \big]\;,
\end{equation}
where we have included the non-dynamical gravitational background.
Note that after integrating over the ``hidden variable'' parameters $\tilde{x}$, we 
obtain an effective theory of visible and dual (dark) matter in the observed spacetime $x$ :
\be
S_\mathrm{eff} = - \int\! \sqrt{{g(x)}} 
\left[ L_m (A(x)) + \tilde{L}_{dm}   ( \tilde{A}(x))+ \cdots \right]\,.
\label{e:TsSd2}
\ee
Thus, the metaparticle can be understood as a generic message of string theory/quantum gravity for low energy physics.
Like their visible particle cousins, dark matter quanta should be detectable through their particular metaparticle correlation/entanglement to visible matter.
This is a Berry-phase-like effect that comes from a fully covariant description, and is uniquely different
from the usual effective field theory interaction terms between visible and dark matter particles.
We will discuss the observable consequences of this view of dark matter in the last
section.

\section{Quantum gravity, dark matter and dark energy} 


We  now explain how the generalized geometric formulation of string theory discussed above provides for an effective description of dark energy that is consistent with de Sitter spacetime. 
This is essentially due to the theory's chirally and non-commutatively~
doubled realization of the target space,
and the stringy effective action on the doubled non-commutative~
spacetime $(x^a,\tx^a)$ :
\be
S_{\text{eff}}^{\textit{nc}}
  =\iint \text{Tr} \sqrt{g(x,\tx)}\, \big[R(x,\tx) +L_m(x,\tx) +\cdots\big]\;,
\ee
where the ellipses denote higher-order curvature terms induced by string theory, 
and $L_m$ is the matter Lagrangian.
$S_{\text{eff}}^{nc}$ clearly expands into numerous terms with different powers of $\lambda$, which upon $\tx$-integration and from the $x$-space vantage point produce
various effective terms.
To lowest (zeroth) order of the expansion in the non-commutative parameter $\lambda$
of $S_{\text{eff}}^{\textit{nc}}$ takes the form:
\be
S_d = - \iint\! \sqrt{-g(x)} \sqrt{-\tilde{g}(\tx)} \big[R(x) + \tilde{R}(\tx)\big]\;,
\ee
a result which was first obtained almost three decades ago,  effectively neglecting $\w_{AB}$ 
by assuming that $[\hat x^a,\htx_b]=0$.\cite{Tseytlin:1990hn}\ \
In this leading limit, the $\tx$-integration in the first term 
defines the gravitational constant $G$, and in the second term produces a {\it positive} cosmological constant $\L>0$. 
In particular, we are lead to the following low-energy effective action valid at long distances of the observed accelerated universe (focusing on the relevant $3{+}1$-dimensional 
spacetime $X$, of the ${+}\,{-}\,{-}\,{-}$ signature):
\be
S_{\text{eff}} =\frac{-1}{8 \pi G}\int_X \sqrt{-g}
  \big( \L + {\textstyle\frac{1}2} R + {\cal O}(R^2)
\big)\;,
 \label{e:Seff}
\ee
with $\L$ the positive cosmological constant (corresponding to the scale of $10^{-3}$ eV) and
the ${\cal O}(R^2)$ denote higher order corrections (which are also required by the sigma model of string theory~\cite{rF79b}).

It also follows from this construction that the weakness of gravity is determined by the size of the canonically-conjugate dual space, while the smallness of the cosmological constant is given by its curvature. 
Given this action, we may proceed reinterpreting Ref.~\citenum{Tseytlin:1990hn}.
Integrate out the dual spacetime coordinates and write
the effective action as
\begin{equation}
\overline{S} \sim \tilde{V} \int_X\!\! \sqrt{-g(x)} R(x)+ \cdots\;,
\quad\text{where}\quad
\widetilde{V} = \int_{\tilde X}\!\! \sqrt{-\tilde{g}(\tx)}\;,
\end{equation}
and then relate the dual spacetime volume to the observed spacetime volume as
$\widetilde{V} \sim V^{-1}$ (T-duality). This produces 
an ``intensive'' effective action\cite{Tseytlin:1990hn} :
\be
\overline{S} 
= \frac{\int_X\! \sqrt{-g(x)} \left(R(x) +L_m(x)\right)}{ \int_X\! \sqrt{-g(x)}}+\cdots
\;.
\ee
By concentrating on the classical description first (we discuss below 
 quantum corrections and the central role of intrinsic non-commutativity in string theory) we obtain the Einstein equations~\cite{Tseytlin:1990hn}
\begin{equation}
R_{ab} - \frac{1}{2} R g_{ab} +T_{ab} + \frac{1}{2} \overline{S}\, g_{ab} =0\;,\quad\text{with}\quad
T_{ab} \overset{\scriptscriptstyle\text{def}}= \frac{\partial L_m}{\partial g^{ab}} - \frac{1}{2} L_m\,g_{ab}\;.
\end{equation}
We emphasize that our reinterpretation of Ref.~\citenum{Tseytlin:1990hn}
does not follow the original presentation and intention.
In particular, we directly relate the intensive action to the cosmological constant, 
$\overline{S}\sim \Lambda$. 
Note that this new approach to the question of dark energy (viewed as a cosmological constant) in quantum gravity
is realized in certain stringy-cosmic-string-like~
toy models,\cite{rBHM7} 
which can be viewed as illustrative of a generic non-commutative phase of F-theory.\cite{rFTh}
In particular, a ``see-saw" formula is directly realized in Ref.~\citenum{rBHM7} 
as $M_\L\,{\sim}\,M^2/M_P$,
where $M_\L$ is the dark energy scale, $M_P$ the Planck scale, 
and $M$ an intermediate scale coming from the matter sector (such as the Higgs scale).

Note that, in general, 
to lowest (zeroth) order of the expansion in the non-commutative parameter $\lambda$,
$S_{\text{eff}}^{\textit{nc}}$ takes the following form (that also includes the matter sector and its dual) \cite{BHM} :
\be
S_d = - \iint\! \sqrt{{g(x)}{\tilde{g}(\tx)}} 
\left[R(x) + \tilde{R}(\tx)+ L_m (A(x, \tilde{x})) + \tilde{L}_{dm}   ( \tilde{A}(x, \tilde{x})) \right].
\label{e:TsSd}
\ee
Here, $A$ denotes the usual SM fields, and $\tilde{A}$ denotes their duals.
Note that after integrating over the dual spacetime, and after taking into account T-duality, the intensive action now reads
\be
\overline{S} 
= \frac{\int_X\! \sqrt{-g(x)} \big(R(x) +L_m(x) + \tilde{L}_{dm} (x)\big)}{ \int_X\! \sqrt{-g(x)}}+\cdots\;.
\label{e:TsLb}
\ee
The proposal here is that the dual sector (as already indicated in the previous section) should be interpreted as the dark matter sector,
which is correlated to the visible sector via the dark energy sector, as discussed in Ref.~\citenum{Ho:2010ca}. 
(The radiative stability of this construction has been discussed in Ref.~\citenum{BHM}, which also addresses the hierarchy problem.)
We emphasize the unity of the description of the entire dark sector based on 
the properties of the dual spacetime.
Note that one can also discuss statistical effects in this view of dark energy and,
in particular, provide an explicit formula for a dynamical form of dark energy
\cite{Jejjala:2007hh} that can be compared to cosmological observations.


\section{Coda: Quantum gravity phenomenology and the real world} 

In conclusion, we discuss the implications of our new view on dark matter and dark energy
in the context of actual astronomical observations.

\subsection{Observational Issues}

The above metaparticle-like dark matter quanta are by construction correlated to visible matter and have been discussed in the literature as
Modified dark matter (MDM).\cite{Ho:2010ca}  
MDM is, at the moment, a phenomenological model of dark matter inspired by gravitational thermodynamics. For an accelerating Universe with positive cosmological constant $\Lambda$, certain phenomenological considerations lead to the emergence of a critical acceleration parameter related to $\Lambda$.
This ``fundamental acceleration'' is just the value of $\Lambda$ expressed as acceleration $\sim$$\,$$cH_0$, where $H_0$ is the Hubble constant, and thus, is of the order of $10^{-10}\mathrm{m/s^2}$. 
Appearance of this acceleration scale in the data is an expected manifestation of MDM, and 
its existence is observationally supported as discussed below.
%
The resulting MDM mass profiles, which are sensitive to $\Lambda$, are consistent with observational data at both the galactic and cluster scales. In particular, the same critical acceleration appears both in the galactic and cluster data fits based on MDM.\cite{Ho:2010ca}
Furthermore, using some robust qualitative arguments, MDM appears to work well on cosmological scales.
If the quanta of MDM are metaparticles,
this may  explain why, so far, dark matter detection experiments have failed to detect dark matter particles. In particular, the natural model for MDM quanta could be
provided by the metaparticle realizations of the SM particles, associated with bi-local extensions
of all SM fields. Thus the baryonic matter described by the SM fields (the $A$ backgrounds in the
above discussion) would have natural
cousins (the $\tilde{A}$ backgrounds in the above discussion) in the dark matter sector, which in turn would be sensitive to the dark energy modeled by the cosmological
constant $\Lambda \sim 1/H_0^2$, that is, the curvature of the dual spacetime,
which is radiatively stable and related to the Planck energy and the characteristic energy scale of the visible sector $M_\L\,{\sim}\,M^2/M_P$.

The importance of the fundamental acceleration $10^{-10}\mathrm{m/s^2}$ 
is manifest in the empirically established
baryonic Tully-Fisher\cite{TFR} (BTF), and
baryonic Faber-Jackson\cite{FJR} (BFJ) relations.
The BTF relation refers to the observed correlation between 
the total baryonic mass of spiral galaxies $M_\mathrm{bar}$
and the rotational velocity in the flat part of the rotation curve $v_{\textrm{flat}}$ of the form
$M_\mathrm{bar} \sim v_{\textrm{flat}}^4$. 
The slope of the BTF relation gives the fundamental acceleration
$v_{\textrm{flat}}^4/(G M_\mathrm{bar}) \sim 10^{-10}\mathrm{m/s^2}$.
%
%
The relation for elliptical galaxies and other pressure supported systems that parallels
the BTF for spirals is the BFJ relation 
between $M_\mathrm{bar}$ and
the line-of-sight velocity dispersion $\sigma$ of the form $M_\mathrm{bar} \sim \sigma^n$.
When fit to data of various pressure supported systems, 
the best fit to the power $n$ was found to be anywhere from about 3 to 5
depending on the scale of the systems considered. 
However, when the data for globular clusters, elliptical galaxies, galaxy clusters, etc. are
all analyzed together, the preferred power is $n=4$, though the data does show
considerable scatter around this line.\cite{Farouki1983,Sanders1994,EMT2021}
Furthermore, the slope of the $n=4$ BFJ fit again gives
the fundamental acceleration 
$\sigma^4/(G M_\mathrm{bar}) \sim 10^{-10}\mathrm{m/s^2}$.

According to the $\Lambda$CDM model of cosmology, our Universe started out with almost
uniform distributions of cold dark matter (CDM) and baryonic matter.
Consequently, a correlation between the total dark matter and total
baryonic matter in a region surrounding a galaxy is to be expected in the $\Lambda$CDM model.
However, given that galaxies are thought to have gone through various phases including starbursts,
emission of gasses, and multiple mergers during their evolutionary histories, a dynamic
correlation such as the BTF and BFJ relations involving the fundamental acceleration scale is surprising. 
%
The question is: what do these correlations imply on the nature of dark matter?
What type of dark matter would be able to explain these relations?

\subsection{Analogy with Turbulence}

Given that the temperature fluctuations in the Cosmic Microwave Background have
been observed to be scale invariant,\cite{Planck:2018vyg}
and assuming that dark matter (whether traditional or metaparticle) is collisionless, then
an analogy with Kolmogorov's theory of turbulence may be instructive.\cite{Frisch}

In the so-called inertial range of the turbulent fluid, \textit{i.e.} at length scales $L$ where the
viscosity of the fluid can be neglected, the dynamics of vortices is determined solely by 
the rate of energy dissipation $\varepsilon$, and if one writes
$\vev{\mathrm{KE}} = \int dk\,E(k)$, where $\vev{\mathrm{KE}}$ is the mean turbulent kinetic energy
of the flow and $k=2\pi/L$ is the wave number, simple dimensional analysis
yields the famous Kolmogorov scaling relation $E(k) \propto \varepsilon^{2/3} k^{-5/3}$, which has
been confirmed experimentally.
The length scale $\eta$ below which viscous effects become important, called the Kolmogorov microscale, 
is the boundary of the inertial region.
Again, from dimensional analysis, it is argued that $\eta \propto (\nu^3/\varepsilon)^{1/4}$,
where $\nu$ is the kinematic viscosity.
Essentially, the scale $\eta$ is determined by the competition between $\varepsilon$ and $\nu$.
Note that this argument assumes that there is no other dimension-ful scale in the problem other
than these two parameters.

In the case of structure formation in a statistically self-similar (\textit{i.e.} scale invariant) Universe,
the two competing effects would be attraction due to gravity, parametrized by $G$, and the speed of the 
expansion of the Universe $H(z) = \dot{a}/a$, where $z$ is the redshift.
The scale of the structures that form at redshift $z$ should be determined by $G$ and $H(z)$.
Now, consider the size $R_\mathrm{vir}$ of a virialized structure of total mass $M$.
We expect $\vev{v^2}\sim \sigma^2 \sim GM/R_\mathrm{vir}$.
The gravitational acceleration at the edge of this structure is
$a_\mathrm{vir} = GM/R_\mathrm{vir}^2 \sim \sigma^4/(GM)$.
Thus, the scale of the structure $R_\mathrm{vir}$ 
can be characterized by this acceleration scale $a_\mathrm{vir}$.
From dimensional analysis a la Kolmogorov we expect
$a_\mathrm{vir} \sim \sigma^4/(GM) \propto cH(z)$,
which is analogous to the $n=4$ FJ relation except
1) $M$ is the sum of the dark matter mass and the baryonic mass, 
2) dark matter halos are not necessarily virialized,
3) the proportionality constant may not be order one,
4) the $z$ dependence of $H(z)$ suggests that $\sigma^4/(GM)$ will depend on 
the redshift $z$ at which the structures formed, and
5) in actuality, there are many other scales present.
Nevertheless, this simple handwaving argument does suggest
a possible path toward the BFJ, or a BFJ-like relation.
Note also that metaparticle quanta
are sensitive to both the UV ($G$, the Planck scale) and the IR ($H(z)$),
suggesting that they may provide the linchpin connecting those scales,
and the crucial ingredient in constructing a dark matter model 
that could realize this scenario.

One intriguing prediction of the above discussion is the possible $z$-dependence of
$\sigma^4/(GM)$.
Replacing $M$ by $M_\mathrm{bar}$ should not erase this dependence.
Dimensionally, $\sigma^4/(GM_\mathrm{bar})$ should scale as $(1+z)$.
Galaxy clusters have formed recently $(z<1)$,
galaxies have started to form at around $z\sim 10$, and
globular clusters at around $z\sim 10^2$.\cite{Bothun1998}
So if $\sigma^4/(GM_\mathrm{bar})$ is around $10^{-10}\,\mathrm{m/s^2}$ for galaxy
clusters, it should be around $10^{-9}\,\mathrm{m/s^2}$ for galaxies,
and $10^{-8}\,\mathrm{m/s^2}$ for globular clusters.
In Fig.~\ref{fig:snowman} we plot $\sigma^4/(GM_\mathrm{bar})$
for a variety of structures.
As can be seen, the galaxy cluster data points are clustered around
$10^{-10}\,\mathrm{m/s^2}$, while the galaxy data points are spread out
in both directions and go up as far as $\sim\!10^{-9}\,\mathrm{m/s^2}$, and
the globular cluster data are spread out even further and go up as far as 
$\sim\!10^{-8}\,\mathrm{m/s^2}$.
One interpretation of this result is that
galaxies and globular clusters started out with characteristic accelerations of
$O(10^{-9}\,\mathrm{m/s^2})$ and $O(10^{-8}\,\mathrm{m/s^2})$, respectively,
but have migrated to lower values as they went through various stages of evolution,
resulting in distributions centered around $10^{-10}\,\mathrm{m/s^2}$ (in the log scale).
This possibility had not been manifest in previous BFJ analyses
which had always plotted $\sigma^4$ against $M_\mathrm{bar}$,
demonstrating that a simple change of perspective can open up new vistas.

\begin{figure}[ht]
\begin{center}
\includegraphics[width=9cm, angle=-90]{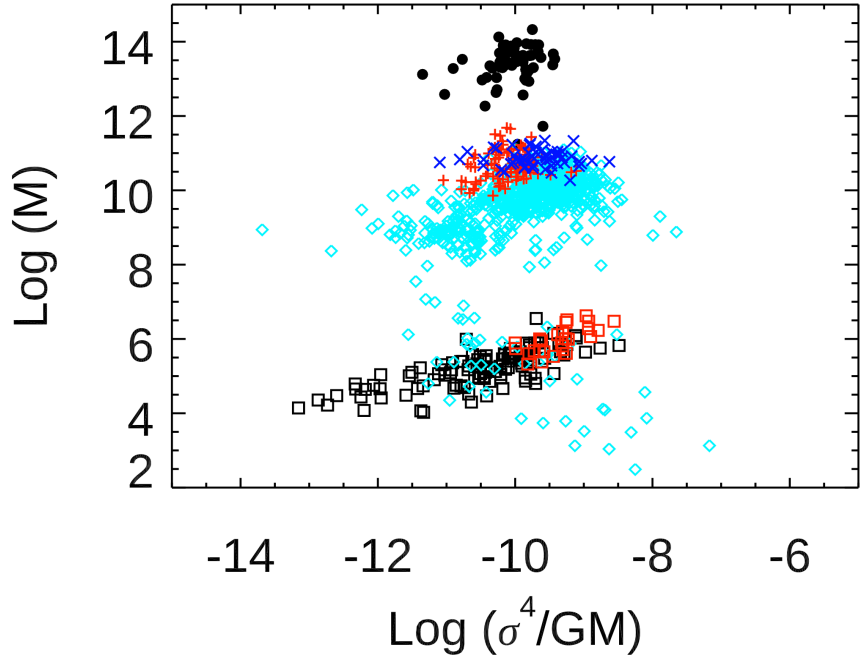}
\end{center}
\caption{
{The virial acceleration for structures spanning about 10 decades of scale $M/M_\odot = 10^{4\sim 14}$. 
The data points are: $\bullet$ galaxy clusters \cite{Zhang:2010qk},
{\scalebox{0.8}{$\red{\bm{+}}$}} elliptical galaxies \cite{Cappellari:2012ad,Cappellari:2012ae},
{\scalebox{0.8}{$\blue{\bm{\times}}$}} elliptical galaxies \cite{Belli:2013oia,Belli:2014yfa},
$\cyan{\bm{\diamond}}$ elliptical, dwarf elliptical, and dwarf spheroidal galaxies \cite{stw1248},
{\scalebox{0.7}{$\bm{\square}$}} Milky Way globular clusters \cite{Baumgardt:2018},
{\scalebox{0.7}{$\red{\bm{\square}}$}} M31 (Andromeda) globular clusters \cite{Strader:2009hg}.}
}
\label{fig:snowman}
\end{figure}

\subsection{Closing Remarks}

In this paper, we propose that the BFJ relation could be explained by metaparticle-based, or similar MDM models.
We are, however, mindful that it could well be explained by more conventional means.
Indeed, Kaplinghat and Turner \cite{KT2002} have provided a scenario on how the BTF relation may emerge from $\Lambda$CDM,
though we are unaware of a similar work on the BFJ relation.
MDM models, though suggestive, 
are lacking in similar concrete scenarios that would connect the models to the BFJ relation;
a direction of research we intend to pursue.

This concludes our presentation of the new approach to quantum gravity (based on modular spacetime, Born geometry and metastring theory) and quantum gravity phenomenology in the context of dark matter and dark energy and actual astronomical observations.

\medskip
\noindent
{\bf Acknowledgements:}
We are grateful to L. Freidel, R. G. Leigh, J. Kowalski-Glikman, 
P. Berglund, T. H{\"u}bsch, D. C. Dai, D. Stojkovic, J. H. Simonetti, M. Kavic and V. Jejjala
for discussions. 
The research of DM and TT is supported in part by the US Department of Energy (DE-SC0020262). DM also thanks the Julian Schwinger Foundation, and TT the NSF (PHY-1413031) for support.

%
%
\begingroup
\frenchspacing\raggedright

\endgroup

\end{document}